\documentclass[12pt]{article}
\usepackage{latexsym}
\usepackage{amssymb}
\usepackage{amsmath}
\usepackage{graphicx}
\usepackage{bbm}
\usepackage{enumitem}
\usepackage{slashed}
\usepackage{hyperref}
\usepackage{xcolor}
\usepackage[titletoc,title]{appendix}

\def\be{\begin{equation}}
\def\ee{\end{equation}}

\newcommand{\sm}[1]{\mbox{\scriptsize #1}}
\newcommand{\tn}[1]{\mbox{\tiny #1}}
\renewcommand{\@}[1]{\sqrt{#1}}
\renewcommand{\le}[1]{\label{#1}\end{eqnarray}}
\newcommand{\bea}{\begin{eqnarray}}
\newcommand{\eea}{\end{eqnarray}}
\newcommand{\eq}[1]{(\ref{#1})}

\begin{document}
\pagestyle{plain}

\centerline{{\Large \bf Holography as an Information Principle in Quantum Gravity}}
\vskip.7cm

\begin{center}
{\large Sebastian De Haro$^{1,2,3}$, Enrico Cinti$^{1,2}$, and Aude Corbeel$^2$}\\
\vskip .7truecm
{\it $^1$Institute for Logic, Language and Computation, University of Amsterdam}\footnote{Forthcoming in the \textit{Festschrift in honour of Gerard 't Hooft} on his 80th birthday (World Scientific).}
\\
{\it $^2$Institute of Physics, University of Amsterdam}\\
{\it $^3$Qusoft Research Center for Quantum Software, The Netherlands}\\

\end{center}

\vskip .7truecm

\begin{center}
\today
\end{center}

\vskip 5.5truecm

\begin{center}
\textbf{\large \bf Abstract}
\end{center}

We draw on 't Hooft’s seminal formulation of the holographic principle to analyse the methodological and conceptual role of information in quantum gravity. We argue that, in ’t Hooft’s work and in later developments, information functions as a substantive guiding principle. We distinguish three aspects of this role: First, holographic bounds on the amount of information that can be stored in a region function as theory selection criteria that constrain viable quantum gravity theories. Second, holography functions as a principle of theoretical equivalence: the bulk and boundary theories must describe the same physical content, even though ’t Hooft privileges a more fundamental, lower-dimensional, and potentially deterministic boundary description. Third, the distribution and encoding of information connect ’t Hooft’s proposals to contemporary work on bulk reconstruction, holographic quantum error correction, and ER=EPR, where emergent spacetime structure is tied to patterns of entanglement and redundancy. On the basis of these three roles, we argue that purely epistemic or Shannon-style conceptions of information are inadequate in this context: we instead outline a distinction between what we call maximal and intermediate conceptions that aim to capture the methodological and interpretative roles of information in holographic quantum gravity. We therefore suggest that, in the context of holography and quantum gravity, a more systematic philosophical treatment of the role of information as an interpretation-guiding principle would be desirable.

\newpage

\tableofcontents

\section{Introduction}\label{intro}

It is an honour to contribute to the {\it Festschrift} for Gerard 't Hooft's 80th birthday. Our contribution will focus on his seminal work on the holographic principle, `Dimensional reduction in quantum gravity' ('t Hooft, 1994), a paper written for the {\it Salamfestschrift} that first appeared as a preprint in 1993. We will argue that, in this paper and in subsequent work in quantum gravity by other authors, information content plays the role of a guiding physical principle. Specifically, entropy bounds and microstate counting arguments constrain what are considered to be admissible theories of quantum gravity. We distinguish three roles that information plays in 't Hooft's work, and argue that these roles illustrate how information, understood in a statistical-mechanical sense, has a methodologically central guiding function in quantum gravity practice. Furthermore, this function sits in tension with some of the standard analyses in philosophy of physics, which treat information as epistemic, and largely dissociated from questions of semantics and interpretation. 

The holographic principle has proven to be highly fruitful in quantum gravity.\footnote{The holographic principle was first formulated in 't Hooft (1994) and further developed in Susskind (1995). Since our aim here is not to give an overview of the field, and is also not historical, but rather to discuss some conceptual issues related to 't Hooft (1994), we will restrict the number of references to the original works on the subject, not least because they are very numerous. For reviews of the holographic principle, see 't Hooft (2001) and Bousso (2002).} 
In particular, the AdS-CFT correspondence has turned holography into a precise duality conjecture between a gravitational theory in the bulk and a non‑gravitational quantum field theory on the boundary. This has generated results on a wide range of topics, such as black hole entropy, renormalisation, and strongly coupled gauge dynamics, and has inspired further work on holography outside the original AdS-CFT setting.\footnote{For reviews of AdS-CFT, see e.g.~Aharony et al.~(2000) and Ammon and Ermenger (2015). For a philosophical review, see De Haro, Mayerson and Butterfield (2016).} 
In what follows, we use 't Hooft's early formulation as a lens through which to view some of these later developments, with a focus on the role of information content in making holographic ideas methodologically and physically productive.

In philosophy of physics, information is often treated in an epistemic or in a purely formal way. Thus it is often taken to concern either (1) agents' beliefs and knowledge, or (2) formal systems such as collections of uninterpreted data or syntactic structures, rather than itself playing a substantive role in the physics and its interpretation. Thus information is often related to the compressibility of data and to uncertainty. For instance, some authors interpret Shannon entropy as a measure of the expected gain in knowledge (or reduction in uncertainty) that an agent would obtain upon receiving a message, and they use this interpretation to criticise the argument used by Bekenstein to attribute thermodynamic entropy to black holes (Dougherty and Callender, 2017; Wüthrich, 2017). By contrast, Timpson (2013:~pp.~4, 25) argues that Shannon entropy is a measure of the compressibility of an information source, and should be distinguished from the everyday notion of information associated with knowledge or meaning.\footnote{In other fields like philosophical logic, information is also a central topic. There are various philosophical accounts of information, including those that relate information to data and computation (Shannon, 1948; Kolmogorov, 1965), that treat information as syntactic, and that define information as semantic and factive (Sequoiah-Grayson and Floridi, 2022; Dretske, 1981; Floridi, 2011). For overviews, see Adriaans (2023), Cover and Thomas (2006), and Adriaans and van Benthem (2008).\label{inforefs}}
(We will refer to theses (1) and (2) collectively as `Shannon-syle' accounts of information.)\footnote{Timpson's view also draws on the extension of Shannon's `source coding theorem' to the quantum realm by Schumacher (1995).}

There are indeed good reasons not to treat information as a thing or as a new type of physical entity, and to stay away from popular glosses like `It from bit' or any supposed easy guide to fundamental ontology. As Timpson (2013:~pp.~5, 70--72) argues, Wheeler's (1989:~pp.~310-311) slogan ‘It from bit' can be read as expressing a form of informational immaterialism, according to which material entities such as particles and fields are ultimately replaced by an underlying basis of information. The rejection of such a view, together with the rejection of any radical Pythagoreanism according to which the physical world is fundamentally mathematical, helps to secure a distinction between physical systems and mathematical structures. Furthermore, treating information as a concrete spatiotemporal entity whose trajectory must be tracked has led to well known conceptual puzzles, particularly in attempts to understand how information is transmitted in quantum teleportation (Timpson:~2013, p.~83).

Despite these caveats, treating information as being, in effect, irrelevant for the interpretation of physical theories, also runs the risk of throwing out the baby with the bathwater. For it has been a fruitful standard practice in quantum gravity to take information as a {\it guiding physical principle} in developing new theories. A pressing  question, then, is in what ways arguments based on information content can constrain and shape physical theories. 

To make this more concrete, we focus on cases where information content is used as a constraint on the admissible state-space, quantities, and dynamics of a theory.

Thus our main observation will be that, to a large degree, holography has been fruitful because information-based considerations of the sort already prominent in 't Hooft (1994) are taken seriously as guiding principles that constrain theory space and shape proposals for dualities and emergent spacetime. In this sense, holography functions as an information principle: it constrains the amount, equivalence, and distribution of physically admissible information in quantum gravity theories. To illustrate this, we will distinguish three aspects of the notion of information in 't Hooft (1994) that we will compare to more recent ideas in quantum gravity:

{\bf (i) The amount of information:} Entropic arguments and the amount of information are argued to constrain physically viable theories of quantum gravity.

{\bf (ii) Theoretical equivalence:} Holography is understood a stating an equivalence of descriptions, including the bulk and the boundary theories describing the same physical systems, states, quantities, and dynamics.

{\bf (iii) The distribution and encoding of information:} Non‑local EPR-type correlations and the possibility of decoding bulk information from boundary data (as in quantum error-correction and ER=EPR-type proposals) are taken to be physically significant for how information is distributed in spacetime and to constrain physically viable models of quantum gravity.

The next three Sections develop these three points in sequence. We will argue that, taken together, they point to a mismatch between the fruitful substantive role that is given to the notion of information in quantum gravity practice, and the caution, in some discussions in the philosophy of physics on black holes and Shannon entropy, about whether such roles should be assigned to information in physics. By `substantive', we here mean roles that go beyond both the beliefs of agents and the mere counting uninterpreted data, i.e.~along the lines of points (i) to (iii). Our goal here is not to articulate a general framework for thinking about the role of information in physical theories, nor to assess which of the existing accounts of information in the literature (see footnote \ref{inforefs}) does the job best. Rather, our goal is to constrain those projects by providing evidence that, in order to fulfil the three roles that information plays in physical practice, any notion of information must be at least in part interpretative. Thus Section \ref{notionofinfo} will discuss two general conceptions of information, both of which go well beyond Shannon-style accounts and incorporate both structural and interpretative elements.

As a point of terminology, we note that 't Hooft uses both the phrases `dimensional reduction' and `hologram' for what soon after became known as the `holographic principle'. It will therefore not be anachronistic for us to use the phrase `holographic principle' for the main idea of his paper.

The structure of the paper is as follows. Section 2 discusses the amount of information, understood through entropy bounds, and shows how ’t Hooft’s original argument constraining quantum field theories by black hole thermodynamics mirrors more recent Swampland entropy arguments that limit consistent effective field theories. Section 3 discusses holography as a principle for the equivalence of bulk and boundary theories. This Section also points to another idea in ’t Hooft’s paper, regarding the asymmetric stance on fundamentality and his proposal that a deterministic boundary dynamics underlies quantum mechanics. We argue that there is a similar tension between duality and emergence in the more recent literature. Section 4 discusses the distribution of information, and connects ’t Hooft’s idea of infinitely correlated bulk degrees of freedom and emergent bulk descriptions to contemporary discussions of bulk reconstruction, holographic quantum error-correction, and ER=EPR. Section \ref{notionofinfo} discusses two conceptions of information, a maximal and an intermediate one, that both go beyond Shannon-style notions and make explicit the interpretative elements needed for information to play its roles in holographic practice. Section \ref{conclusion} concludes.

\section{The amount of information}\label{firstT}

This section develops aspect (i) from Section \ref{intro}: namely, the amount of information that can be stored in a finite region, given by quantum statistical mechanical entropy, and its role in constraining viable quantum gravity theories. 

\subsection{'t Hooft's entropic argument}

't Hooft's main argument is that entropy bounds can be used to limit the information content of any physically admissible theory of quantum gravity, and he makes this vivid by comparing an ordinary QFT estimate with the entropy of a black hole filling the same region. Thus he considers an ordinary QFT in a spherical region of radius $R$, with a thermal state at temperature $T$, typically a thermal gas of relativistic particles. For such a gas, the energy density grows like $T^4$ while the entropy density grows like $T^3$. So the total energy grows like volume times $T^4$ and the total entropy grows like volume times $T^3$.

But if one tries to excite all those degrees of freedom, one packs so much energy into the region that it would lie inside its own Schwarzschild radius and form a black hole. For as $T$ increases, the energy rises faster than the entropy (because of the power 4 vs.~the power 3 in $T$), so beyond some point the total energy in the region would exceed the mass of a Schwarzschild black hole of radius $R$. This means that, in effect, we are not considering a region of radius $R$, but rather a larger region. Thus if the radius of the region is fixed, one must require that the energy does not exceed the black hole mass. Since the energy depends on the temperature, this gives a maximum physically allowed temperature for the QFT in that region. Putting that maximal temperature into the QFT entropy estimate, one finds that even at this extreme, the QFT entropy grows more slowly (like the 3/4 power of the area) than the entropy of a black hole of the same size, which grows exactly with the area of the horizon.

So, for a region of fixed size, no ordinary QFT configuration can reach the entropy of the corresponding black hole without already having collapsed. That means that, for any closed spacelike surface with the topology of a two-sphere and surface area $A$, the Bekenstein-Hawking black hole entropy is the upper limit, which 
in Planck units is given by:\footnote{See Bekenstein (1973:~p.~2337; 1981:~p.~287). For generalizations of the holographic idea to arbitrary spacetimes, see Bousso's (2002) covariant entropy bound.}
\bea\label{BHentropy}
S_{\tn{BH}} = {A\over4}~.
\eea
Recall that Eq.~\eq{BHentropy}, interpreted as a quantum statistical mechanical entropy, counts states in a microcanonical ensemble at fixed energy.\footnote{For discussions of this, see e.g.~Wallace (2019).} 
Thus this entropy counts the number of mutually orthogonal quantum states that there can be in that volume.\footnote{As we will discuss in Section \ref{de}, 't Hooft ultimately wants to think of the microscopic states as classical states, and of the quantum states as emergent. However, we here endorse the standard holographic interpretation in terms of a dual quantum field theory. We will return to 't Hooft's more speculative idea of classical states in Section \ref{de}.} 
Writing $S_{\tn{BH}} = \ln2^n$ for $n$ Boolean degrees of freedom, one obtains $n = A/(4 \ln 2)$. 't Hooft concludes that the number of physically allowed degrees of freedom in a region must scale with its boundary area, not its volume, which he takes as the core quantitative hint from which to formulate the holographic principle. Ordinary QFT overcounts states unless, through gravity, its Hilbert space is reduced to obey this area law.

Section \ref{secondT} will discuss the holographic interpretation of this argument in more detail. Suffice it here to say that, since this is a very general argument about the allowed energy in a region in the presence of gravity, and about the area-dependence of the corresponding degrees of freedom, 't Hooft's bound is taken to be a {\it selection principle} on physically possible theories of quantum gravity. An ordinary local QFT with a Planckian cutoff has too many states in a given region: most of them would correspond to configurations that already lie inside their own Schwarzschild radius and so must actually be described as black holes in a larger region. Any viable quantum gravity theory must therefore reduce its Hilbert space so that, for each region, the dimension grows only according to the area law.

This illustrates the main theme of this Section: the amount of information which is given by the quantum statistical-mechanical entropy functions as a guiding principle, though not in the epistemic sense of `our knowledge of the system'. Rather, information here plays the role of a physical constraint on the kinds of states that, in the presence of gravity, can be realised (see Section \ref{secondT}). Thus construed, black hole entropy measures the information capacity of a finite region: namely, the logarithm of the number of physically realisable microstates compatible with its macroscopic parameters. By contrast, if entropy were given a purely epistemic construal, as quantifying agents’ uncertainty, the area law would at best constrain what we can know about a region: it could not justify ’t Hooft’s use of the entropy bound as a physical selection principle on admissible Hilbert spaces: it would not suggest a holographic principle.\footnote{For more discussion on how the concept of information is used in black hole physics, see Corbeel and De Haro (2026).}

\subsection{Entropic arguments in the Swampland programme}

While discussions of holography are often framed in the context of AdS-CFT, this Section will instead focus on a more recent, and less widely known, illustration of 't Hooft's entropy reasoning that does not involve AdS: namely, the entropy argument in the context of the `swampland programme'. This programme is formulated in terms of effective field theory and makes especially clear how information‑theoretic bounds can function as selection principles in quantum gravity. The example is of interest because its primary aim is not to construct a holographic theory, but rather to constrain possible low-energy theories of quantum gravity. In this way, the example also illustrates the broader influence of 't Hooft's reasoning, beyond strictly holographic scenarios.

The swampland programme addresses the landscape of possible quantum gravity theories (with its wide variety of possible fields, gauge groups, interactions, etc.) by asking which low-energy effective field theories (EFTs) can arise from a UV-complete theory such as string theory, and which lie in the “swampland” of apparently consistent, but ultimately inadmissible, theories (Ooguri and Vafa, 2007). The conjectured criteria for a theory to belong to the landscape include the distance and weak gravity conjectures, which link large distances in field space to the appearance of towers of light states, and require gravity to be the weakest force, respectively. These conjectures are increasingly being understood as following from more basic physical principles such as the finiteness of quantum gravity. (We will not here discuss these conjectures in detail, but rather give the physical argument involving black holes that is parallel to 't Hooft's argument.)

Hamada et al.~(2022:~pp.~6--7) is an example of this shift. Their entropy argument for the distance and weak gravity conjectures uses the same kind of energy and entropy reasoning discussed in the previous Section: they consider putting a large number of species of matter (i.e.~different particle species, corresponding to distinct quantum fields) into a finite region of radius $R$, they estimate the entropy of this gas, and then they require that the total energy not exceed the mass of a black hole of the same size.\footnote{For related arguments, see Dvali and Redi (2008:~p.~4), who discuss black holes in a theory with many species and derive a bound on the temperature, close to 't Hooft's. Their larger argument is that the gravity cutoff is lowered by a power of the number of species, with this number of species bounded, which they propose as a possible resolution of the hierarchy problem. See also Dvali (2010:~p.~532).} 
In their case, the “species” are small black holes, each charged with a distinct type of charge, modelled as pointlike particles, and with masses set by the gauge coupling $g$ and charge $Q$, viz.~$m \simeq g\, Q$. As in 't Hooft's argument, they consider the threshold of gravitational collapse, in a microcanonical ensemble at total fixed energy $E \simeq R$, for a closed surface of radius $R$, and fixed number of particle species, $N_{\tn{species}}$.

The entropy of the gas scales schematically as $N_{\tn{species}}^{1/4}$ times a power of $E$ and $R$, and the allowed energy is bounded by the Schwarzschild mass of a black hole of radius $R$. Requiring that the entropy of the gas never exceeds the Bekenstein-Hawking bound, i.e.~Eq.~\eq{BHentropy}, leads to an upper bound on the number of species, and hence on the charges of the small black holes: $N_{\tn{species}}\simeq Q_{\max}\lesssim R^2$. They argue that a putative violation of this bound, which would follow from the behaviour of the scalar field and the gauge coupling near small extremal black holes, can be avoided if one requires that the EFT cutoff $\Lambda$ is bounded by the coupling: in Planck units, $\Lambda \leq g$.\footnote{They interpret this bound as a magnetic version of the weak gravity conjecture (see Arkani-Hamed et al.~2007). This indicates that there is an infinite tower of light states, so that the EFT loses its validity at large distances.}

Thus the structure of the argument strictly parallels 't Hooft's: in both cases, one starts from an ordinary QFT estimate for a gas in a finite region (relativistic particles for 't Hooft, and small charged black holes treated as particles for Hamada et al.) and calculates how the entropy and the energy vary with temperature, number of species, and size of the region. One then imposes a gravitational consistency condition: the total energy must not exceed the mass of a Schwarzschild black hole fitting in the region, so that the configuration is not larger than the corresponding black hole. This gives the holographic bound on the entropy, Eq.~\eq{BHentropy}. In 't Hooft's case, the conclusion is that an ordinary QFT overcounts states, and the compatibility with gravity requires that the entropy of the Hilbert space scales like the area, rather than the volume. 
In the case of Hamada et al., the same form of reasoning is used to constrain effective theories with gauge and scalar fields. The entropy bound excludes EFTs where one can put arbitrarily many light charged species (i.e.~small black holes) into a fixed region without lowering the cutoff. To avoid having an infinite number of stable small black holes in the model, with the consequent formation of a black hole larger than the volume of the enclosing region, the EFT cutoff must drop with the gauge coupling and, as one goes to infinite distance in field space, an infinite tower of states (i.e.~infinitely many particle species whose masses go to zero, which signals that the EFT is no longer valid) must become light, thus in effect lowering $\Lambda\leq g$ and limiting the size of field space where the theory is valid.

The logic of aspect (i) can be summarised as follows: if a theory allows more states than the Bekenstein-Hawking bound, then it is inconsistent with gravity; so (A) in 't Hooft's case, any QFT with purely volume-scaling entropy is unacceptable unless it is modified, and (B) in the Hamada et al.~swampland case, any EFT with an unbounded number of species at fixed cutoff lies in the swampland. ’t Hooft used this logic to motivate the holographic principle, while Hamada et al.~use it to argue that many apparently consistent EFTs, which ignore the weak gravity conjecture or admit infinite distances in field space without the corresponding towers of light states, cannot arise from quantum gravity: by allowing too many configurations in a finite region, they violate entropy bounds. In both arguments, information content, given quantitatively by the quantum statistical mechanical entropy, plays the role of a theory selection criterion.\footnote{There is a significant difference between the arguments that their different conclusions depend on. In ’t Hooft’s argument the number of species is held fixed and one shows that QFT entropy in a region grows at most with the black hole area law. By contrast, in Hamada et al., the number of species is itself a variable, and the entropy bound is used to constrain how large $N_{\sm{species}}$ can be before the EFT becomes inconsistent with gravity, thus leading directly to the magnetic weak gravity-type cutoff and the requirement that there is a tower of light states that invalidates the EFT.} 
Furthermore, in both cases, entropy is unrelated to agents’ knowledge, data, and computation, and rather sets a physical constraint on which theories and which state-spaces can exist, by using an information-theoretic argument as a selection principle for consistent quantum gravity.

Similar arguments to 't Hooft's (1994) energy and entropy argument and degree of freedom counting have been given in the context of AdS-CFT in Susskind and Witten (1998), who adapt 't Hooft's entropy and area reasoning to anti-de Sitter space by arguing that consistency of the dual CFT again suggests an area-scaling bound on the number of bulk degrees of freedom. Considerations of this type, together with the explicit bulk-boundary dictionary, were the starting point for recognising the holographic character of AdS-CFT.

\section{Theoretical equivalence}\label{secondT}

This section develops aspect (ii) from Section \ref{intro}: holography as a principle of theoretical equivalence. While entropy bounds constrain the number of physically admissible states in quantum gravity theories, holography, for ’t~Hooft, does more than count microstates. It also functions as an equivalence constraint: it requires that the physical content of a bulk description be wholly given by a lower-dimensional boundary description, so that the bulk is, in a sense yet to be clarified, informationally redundant. We also discuss the tension between duality and emergence present both in 't Hooft's work and in recent discussions of dualities.

\subsection{Holography as an equivalence constraint}

't Hooft's main claim can be put as follows. Building on the entropic bound of the previous Section, he argues that `given any closed surface, we can represent all that happens inside it by degrees of freedom on this surface itself' (p.~289). In particular, `one Boolean variable per Planckian surface element should suffice'. He proposes that `there simply {\it are} not more degrees of freedom to talk about than the ones one can draw on a surface' (p.~289). Thus the area law is not only an upper bound on entropy: it is taken to suggest an exhaustive specification of the physical degrees of freedom associated with the interior region in terms of boundary variables, which 't Hooft goes on to clarify.

On this view, a specification of the boundary degrees of freedom gives a complete description of the interior: there are no additional bulk variables that contribute further independent physical information. Any bulk description is therefore informationally redundant, relative to the boundary description. We read this as a constraint akin to requiring a {\it holographic duality}: for any physically possible quantum gravity bulk theory, there must exist a lower-dimensional description whose Hilbert space is isomorphic to the space of physically permitted bulk states. The constraint is not only about the number of degrees of freedom, but also about how these degrees of freedom are encoded on the boundary and in the bulk. Furthermore, the emphasis on {\it are} in the claim that `there simply {\it are}' no further degrees of freedom suggests an ontological identification of the bulk and boundary degrees of freedom: whatever degrees of freedom are being described by the bulk and boundary variables, they are the same degrees of freedom.

On an information-theoretic reading, this isomorphism requirement says that the bulk and boundary descriptions encode the {\it same physical information}: the mapping between them is information-preserving and invertible on the spaces of physically admissible states and sets of quantities. Thus holography becomes a criterion for physically possible quantum gravity theories: bulk theories that do not admit such a reduction are, according to the argument, not physically possible in 't~Hooft's precise sense. We will call this 't~Hooft's {\it viability} criterion for theories of quantum gravity. Section \ref{thirdT} will further develop this information-theoretic reading. 

This interpretation connects naturally to discussions of duality as an isomorphism of state-spaces, quantities, and dynamics. About states: the map from configurations of boundary variables to configurations of bulk variables is injective, because different boundary configurations correspond to different bulk configurations; it is surjective onto the physically possible bulk configurations, because every physically possible bulk state is determined by some boundary configuration. About quantities: 't~Hooft explicitly wants models in which `the data on a two-dimensional surface will determine all observables elsewhere' (p.~291). In his lattice toy model, he requires that `the values of $f$ on a sheet should fix the values elsewhere' (p.~291). Indeed, other work by 't~Hooft (1991:~pp.~250--251) cited in the paper has focussed on holographic algebras as the primary objects from which the theory is to be constructed. Furthermore, he wants the dynamics to match: given an evolution law for the boundary variables, the induced evolution of the bulk variables should reproduce the physics we see. Taken together, these conditions amount to an isomorphism between bulk and boundary descriptions, in the sense relevant to both physics and philosophical discussions of dualities.\footnote{For this conception of duality as an isomorphism of states and quantities that is equivariant for the dynamics, see De Haro and Butterfield (2025:~p.~72).}

In this argument, information again plays a crucial role. A key summary of the paper states that a two-dimensional surface `can contain all information concerning the entire three-space' (p.~290); similar remarks appear in his discussion of the black hole information paradox (p.~286) and of the hologram analogy (p.~289). Thus 't Hooft's holography is not just a reduction of the dimensionality of spacetime, but a more substantive claim that the boundary carries all the physical information about the bulk. It is furthermore naturally read as a case of theoretical equivalence between bulk and boundary theories: the two descriptions must be equivalent in their physical content.\footnote{Part of the philosophical literature on this topic has focussed on whether duals can ``automatically'' always be considered to be theoretically equivalent. The consensus, which we endorse, is that the answer is `No, not always'. A well-known example that illustrates this consensus is the Kramers-Wannier duality of the Ising model: clearly, a hot lattice is not the same as a cold lattice. Thus there are additional conditions for duals to describe the same physical situations. Our own view is that, under these additional conditions, holography as discussed by 't Hooft is indeed a case of physical equivalence. For a summary of the discussion, see De Haro and Butterfield (2025:~Section 12.1). See also De Haro and Cinti (2026:~Section 5.1), Read and M\o ller-Nielsen (2020), and De Haro (2021).} 
In this sense, bulk and boundary provide dual descriptions with the same physical content, even though, as we will see in Section \ref{de}, ’t~Hooft himself interprets the boundary as more fundamental. This is clear from 't Hooft's statements that there {\it are} no more degrees of freedom than those that one can encode on the boundary, and that the boundary contains all the information about the whole three-space.

\subsection{Duality and emergence}\label{de}

An important point about the conception of duality at play in 't Hooft's discussion is that he does not appear to be neutral between the bulk and boundary. This can already transpire from the quotes in the previous section, concerning e.g.~that `given any closed surface, we can represent all that happens inside it by degrees of freedom on this surface itself' (p.~289). This characterization suggests that the physics on the lower-dimensional closed surface should be more fundamental than that happening inside of it, if only because it gives a less redundant description. 

't Hooft however provides a second, more speculative, reason for privileging the boundary in this way: the boundary theory, contrary to the bulk theory, might allow for a completely deterministic reconstruction of quantum mechanics, hence for a deterministic explanation of e.g.~entanglement correlations in the bulk. By `deterministic', 't Hooft here means what is often characterised as a hidden-variable completion of quantum mechanics: a deterministic theory with additional variables, from which quantum mechanics is obtained by a coarse graining procedure over the detailed states of these extra variables. In other words, a theory that stands to quantum mechanics in the same relation in which statistical mechanics stands to thermodynamics.  

The most discussed example of such a theory is the de Broglie-Bohm pilot wave theory (Goldstein, 2025).\footnote{Another example of such a theory are superdeterministic hidden variable theories (Palmer, 2026) or modal intepretations (Van Fraassen, 1991), which, according to Wallace (2008), also count as hidden variable theories. }
't Hooft however discusses a different approach, based on a deterministic cellular automaton whose dynamics is supposed to reproduce quantum mechanics upon coarse graining. The theory in question describes the deterministic evolution of a collection of discrete cells, which aims at providing a completely deterministic underpinning to quantum mechanics, in particular one where quantum probabilities emerge from a deterministic dynamics as a result of our ignorance of the details of the underlying discrete microphysics. As we mentioned, this picture is analogous to the one obtaining between thermodynamics and statistical mechanics. In what follows we remain neutral on the viability of this deterministic completion of quantum mechanics, and focus on how the information-theoretic aspects of 't~Hooft’s proposal constrain admissible theories and motivate holography.

Indeed, 't Hooft goes further and connects this picture to holography. In particular, for 't Hooft, the bulk contains an over-parametrised description of the underlying fundamental physics. This over-parametrisation stems from the inclusion of gravity into our theory. One way to see this is by noting that gravity implies that local degrees of freedom are not gauge invariant, since they are not invariant under diffeomorphisms. So for example, two states differing only by some local excitation in the middle of the bulk should be gauge equivalent.\footnote{There are ways to express the difference between two states of this form, involving dressing of gravitational observables to make them gauge invariant. This, in effect, makes them non-local. In the main text, we are assuming that this procedure is not carried out, and so that we are dealing with genuine local observables.} Since however gravitational theories do involve all kinds of spacetimes differing by the presence of some local excitation, or more generally differing in terms of their local degrees of freedom, this entails that in a gravitational theory gauge invariance forces a significant reduction of the number of true degrees of freedom from the seemingly real, but ultimately gauge, degrees of freedom that we seem able to define. 

't Hooft suggests that, if we remove this redundancy in the bulk gravitational degrees of freedom, we are left with the physics on the closed surface enclosing our gravitational region as the fundamental, microscopic description of the semiclassical, redundant bulk physics. And he suggests, in light of this, that an interesting possibility would be that the physics of this close surface would be that of a deterministic cellular automaton. Indeed, since the arguments from entropy bounds already suggest that we are dealing with a finite number of discrete degrees of freedom on the boundary, the jump to the cellular automaton picture appears more natural in the boundary theory. Moreover, since quantum physics is supposed to emerge from the cellular automaton, this would imply that the bulk physics emerges from the boundary. Indeed, if realized, this idea would lead to a picture where the redundant, bulk quantum theory of gravity emerges from the gauge-invariant, deterministic cellular automaton on the boundary.

When it comes to holography, however, the most interesting consequence of this picture concerns the relation between emergence and dualities. This topic has received significant attention in the philosophical literature, and has led to broad agreement on the following point: duality, being an equivalence relation, hence symmetric, is incompatible with emergence, which is necessarily asymmetric.\footnote{For this formulation of the argument, see De Haro (2017:~p.~118). The argument was given earlier, in a different form, in Dieks et al.~(2015:~p.~209). See De Haro and Butterfield (2025:~Section 14.3) and De Haro and Cinti (2026:~Section 5.3) for discussion and review of these issues.} 
The asymmetry of emergence can be seen, for example, by considering the aforementioned connection between thermodynamics and statistical mechanics, a paradigmatic case of emergence. If thermodynamics emerges from statistical mechanics, then it cannot be the case that statistical mechanics also emerges from thermodynamics: emergence is a one-way street. In the context of holography, this incompatibility between emergence and dualities entails, for example, that we cannot say that the bulk emerges from the boundary, if bulk and boundary are also to be dual.

How is this incompatibility between duality and emergence related to 't~Hooft's idea that the boundary theory, consisting of a deterministic cellular automaton, should be more fundamental than the bulk theory? In one sense, his conclusion is inevitable: if the bulk is claimed to emerge from the boundary, then the boundary must be more fundamental. More interestingly, however, this implies that, in this context, the relation between bulk and boundary is not a standard duality relation: rather, it is an emergence relation between a lower-dimensional theory and a higher-dimensional one. The fact that we can associate the lower-dimensional theory with a boundary and the higher-dimensional one with a bulk is a helpful geometric tool for representing their relation, but does not by itself make them dual in the sense that has become familiar in e.g.~AdS-CFT. A closely related example, already discussed in the philosophical literature (Dieks et al., 2015; De Haro, 2017), is Verlinde's theory of entropic gravity (Verlinde, 2011; Verlinde, 2017). In both cases, including 't~Hooft's proposal, the boundary is ultimately taken to be fundamental and the bulk emergent: an apparently symmetric duality is first assumed at an abstract level, but an additional approximative and interpretative step then introduce an asymmetry and allows for emergence.

\section{The distribution of information: EPR, non-locality, and encoding}\label{thirdT} 

This Section turns to aspect (iii) from Section \ref{intro}: namely, how information is distributed and encoded in holographic theories. We compare ’t Hooft’s ideas to contemporary work on bulk reconstruction, holographic quantum error-correction, and ER=EPR, and thereby discuss both the analogies and the differences present. The guiding idea is that, in these physical scenarios, holography constrains not only how much information a theory may contain, but also how that information is organised, correlated, and redundantly encoded across the bulk and the boundary. 

\subsection{Infinite correlations in three-space and EPR}\label{ic3s}

As we have seen, in 't Hooft's work, the redundancy of bulk degrees of freedom leads to a privileging of the boundary description as the most fundamental one, with the bulk description being understood as emergent. It also leads to speculation on the possibility of a cellular automaton model that might explain, in a deterministic way, the appearance of quantum behaviour. This possibility of using the redundancy of bulk degrees of freedom to explain some apparently puzzling features of quantum mechanics, however, also touches on a further area of contemporary research in holography: bulk reconstruction (Wall, 2014; Dong et al., 2016), quantum error-correction (QEC) (Almheiri et al., 2015), and the ER=EPR conjecture (Maldacena and Susskind, 2013). Before moving to discussing the connection between 't Hooft's work and these areas, we discuss more broadly how 't Hooft thinks about the distribution of information in holographic theories.

As we mentioned, 't Hooft thinks of boundary variables as determining the physics in the bulk, and of the bulk as an overcomplete, redundant representation of boundary microphysics. Indeed, he explicitly states that `physical degrees of freedom in three-space are not independent but, if considered at Planckian scale, they must be infinitely correlated ... The infinite correlations in three-space could also present a new starting point for resolving the Einstein-Rosen-Podolsky paradox.' (p.~290). This suggestion appears to be in line with the idea that the bulk is an overcomplete encoding of a finite collection of boundary bits, in line with 't Hooft's further ideas about cellular automata, and more broadly with his lattice toy model for the boundary theory---and we here remain neutral about these further ideas.

According to the above quote by 't Hooft, the redundancy of the bulk gravitational theory leads to a description in which certain bits are infinitely correlated. Since two infinitely correlated bits can reasonably be considered to be a single bit, we can equivalently describe this situation as one where two apparently distinct bulk degrees of freedom in fact correspond to a single boundary bit of information. EPR-correlations, then, could be remnant bulk manifestations of this boundary unification of disparate, redundant bulk information.

This explicitly suggests that the redundancy in the bulk description might lead to an explanation of quantum features such as entanglement. In this picture, entanglement would not be a fundamental feature, but rather a by-product of the redundant representation of the information on the boundary by the bulk. Entanglement here reflects how boundary information is distributed across the bulk: it is a manifestation of the global constraints that the boundary theory, through the holographic mapping, imposes on the bulk degrees of freedom. Along the lines of 't Hooft's proposal, we can view it as the correlation arising when the information of a single boundary bit is redundantly encoded in many bulk degrees of freedom.

Information, and in particular the way boundary information is distributed across the bulk, plays a substantial role in this approach: bulk phenomena, including apparently irreducible ones like quantum entanglement, ultimately come from the overcomplete and redundant distribution of the boundary information across the bulk. In line with the discussion about emergence in the previous Section, the bulk gravitational theory and its spacetime emerge from the lower-dimensional boundary theory, and this emergence is governed by holography through the way that information is constrained to be distributed across the bulk.

't~Hooft's idea of infinite correlations between bulk degrees of freedom, and his suggested connection with the EPR paradox, remains speculative. He appears to have something close to an ordinary local field theory in the bulk in mind. In the presence of gravity, such infinite correlations would effectively reduce the number of independent physical degrees of freedom to those that can be supported on the boundary. However, holography does not emerge in quite the way familiar from later AdS-CFT constructions. Still, there is an instructive analogy with another mechanism for reducing the number of effective degrees of freedom, which we discuss in the next Section.

\subsection{Comparison with quantum error-correction and ER=EPR}

Section~\ref{secondT} discussed 't~Hooft's claim that 
bulk field degrees of freedom form an overcomplete and hence informationally redundant description. Section~\ref{de} discussed the idea that, in addition, bulk gravitational degrees of freedom may emerge from boundary degrees of freedom by coarse-graining, so that gravitation is like thermodynamics, and the boundary theory is like statistical mechanics. And Section~\ref{ic3s} discussed 't~Hooft's idea that infinite correlations in three-space may in effect reduce the number of independent degrees of freedom of ordinary QFT, and in this way perhaps offer a novel gravitational perspective on EPR.

Two analogous ideas underlie contemporary discussions of the AdS-CFT correspondence. The first is holographic quantum error-correction: bulk degrees of freedom are treated as logical qubits that are encoded, in a deliberately redundant and overcomplete way, in the physical qubits of the boundary CFT, with bulk localisation (for instance via entanglement wedges) providing different ways of packaging the same boundary information.\footnote{The interpretation of the AdS/CFT correspondence as a quantum error-correcting code has been addressed by Bain (2020).} The second idea is ER=EPR, where the spacetime structure in the bulk is related to the entanglement structure on the boundary, so that patterns of quantum correlation help to determine bulk connectivity and geometry.

\subsubsection{Holographic quantum error-correction}

Within the framework of entanglement wedge reconstruction (Wall, 2014; Dong et al., 2016), all physical quantities in the entanglement wedge $EW(R)$ of a boundary spatial subregion $R$ are represented in the CFT by operators acting only on $R$. Roughly speaking, $EW(R)$ is the region of the bulk spacetime that can be reconstructed from the reduced density matrix of the CFT on $R$. It is then standard to say that the entanglement wedge $EW(R)$ of $R$ is the portion of the bulk that is \emph{encoded} in $R$.\footnote{This is sometimes described as a `subregion-subregion duality' (Hubeny et al.~2007; Bousso et al.~2013): a refinement of AdS-CFT that offers a more fine-grained account of how information is distributed between bulk and boundary.}

The term `encoded' is deliberate. Almheiri et al.~(2015) argue that, from the boundary perspective, the possibility of reconstructing bulk operators within the entanglement wedge arises because bulk information is encoded in the boundary as in a quantum error-correcting code. In holographic quantum error-correction, bulk degrees of freedom correspond to \emph{logical qubits}, while boundary degrees of freedom correspond to \emph{physical qubits}. Logical qubits are the error-protected, information-bearing degrees of freedom, realised by encoding them into specific patterns of many physical qubits: as a result, they can survive the loss or corruption of some of those physical qubits.\footnote{In a quantum computer, the logical qubits do not correspond to single hardware qubits, but to quantum information that has been redundantly distributed across many physical qubits. From the computational perspective, logical qubits, rather than individual hardware qubits, are the effective qubits on which quantum computations are performed.} 
Physical qubits are the underlying microscopic constituents in terms of which the code is realised: they provide the substrate that implements the logical degrees of freedom. In the holographic setup, bulk information is stored nonlocally within a boundary region, mirroring ordinary quantum error-correcting codes, where information is protected by ``delocalising'' it in the entanglement structure of many physical qubits. 

This perspective bears an analogy with 't~Hooft's emphasis on the non-local distribution of information. He suggests that `physical degrees of freedom in three-space are not independent but, if considered at Planckian scales, they must be infinitely correlated' (p.~290). In his discussion, these correlations are formulated in terms of bulk degrees of freedom in three-space. By contrast, in the holographic QEC picture, it is first and foremost the entanglement structure of the {\it boundary} theory that underpins the robustness and non-local encoding of bulk information. Thus, although the details of the two proposals differ, both treat non-local correlations as central to how holographic theories organise information.

\begin{figure}
    \centering
    \includegraphics[scale=0.7]{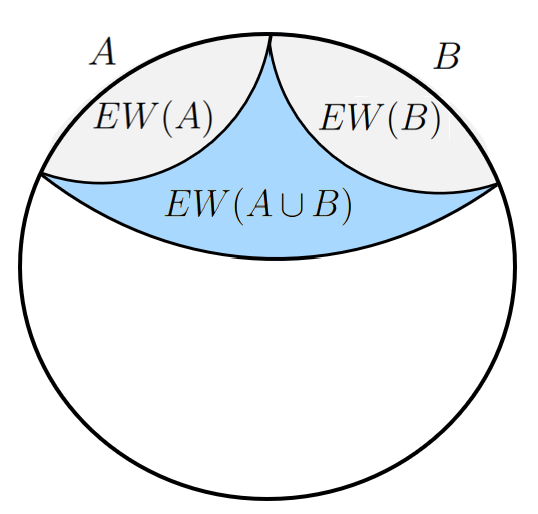}
    \caption{\small Time slice of an AdS geometry illustrating bulk overcompleteness. 
    The entanglement wedges of the individual boundary regions $A$ and $B$, 
    $EW(A)$ and $EW(B)$, are shown in gray, while the larger entanglement wedge 
    of their union, $EW(A \cup B)$, is shown in blue. A bulk operator $\phi(x)$ 
    can lie in the blue region, inside $EW(A \cup B)$ but outside 
    $EW(A) \cup EW(B)$.} 
    \label{ew}
\end{figure}

Furthermore, the analogy can be sharpened in geometric terms. Consider two boundary regions $A$ and $B$. In many states, the entanglement wedge of their union, $EW(A \cup B)$, is strictly larger than the union of the individual wedges, $EW(A) \cup EW(B)$. A bulk operator $\phi(x)$ may lie in $EW(A \cup B)$ while lying in neither $EW(A)$ nor $EW(B)$ separately, see Figure~\ref{ew}. Then $\phi(x)$ cannot be reconstructed from $A$ alone or from $B$ alone, but it can be reconstructed from the combined boundary region $A \cup B$. From the boundary point of view, however, the microscopic degrees of freedom available on $A \cup B$ are the same whether we interpret them as encoding only $EW(A) \cup EW(B)$ or the larger region $EW(A \cup B)$. In this sense, the more fine-grained bulk localisation associated with $EW(A \cup B)$ is an overcomplete way of packaging the same boundary information: relative to the fixed boundary algebra on $A \cup B$, the bulk description exhibits a genuine redundancy in how its local degrees of freedom are assigned. This is a form of {\it bulk} overcompleteness: different localisations of bulk degrees of freedom correspond to the same boundary data, similarly to 't~Hooft's picture, where bulk fields overcount the information present in the boundary theory. 


A different sort of redundancy arises in the boundary encoding itself. The fact that a local bulk field generally admits multiple boundary representations within the CFT, reflects a redundancy in the code: a given bulk observable can be reconstructed from many different boundary subregions, provided they contain its entanglement wedge. Even if part of the boundary is inaccessible or erased, there may remain other subregions from which the bulk information can be recovered. In this sense, bulk information is not only redundantly represented with respect to the boundary as a whole: it is also redundantly encoded with respect to many distinct boundary subregions.

The distinction between logical qubits in the bulk and physical qubits on the boundary, introduced above, illustrates the sense of the emergence of gravity discussed in Section~\ref{de}. Note that, while the logical qubits suffice for a low-energy gravitational description of the bulk, they do not exist independently of the boundary theory: they are realised as redundantly encoded patterns in the physical qubits of the CFT. Thus gravitational dynamics and bulk locality are emergent in this sense: they describe the behaviour of encoded, effective degrees of freedom. This emergence is robust, because the logical qubits are protected: the relevant patterns are non-local on the boundary, and can be reconstructed from multiple boundary regions.

\subsubsection{ER=EPR and entanglement structure}

't~Hooft's emphasis on `infinite correlations in three-space' (p.~290) is also echoed in another important strand of the AdS-CFT dictionary, namely the `ER = EPR' slogan of Maldacena and Susskind~(2013). This conjecture suggests that quantum entanglement (EPR) is associated with geometric connectivity (ER bridges). More recently, it has been given a precise algebraic formulation by Engelhardt and Liu~(2024). The key point is that the \emph{structure} of entanglement (for instance, as classified by the type of von Neumann algebra involved), rather than merely its total amount, is tied to the connectivity properties of the corresponding bulk spacetime. On this view, the \emph{pattern of entanglement encodes aspects of spacetime structure}.

Here again, the details differ from 't~Hooft's proposal. For 't~Hooft, the ultimate aim is to ground quantum behaviour in an underlying deterministic dynamics, with holography playing a crucial role in constraining the admissible degrees of freedom. By contrast, in ER=EPR and related AdS-CFT approaches, the boundary theory is quantum mechanical, and entanglement is not eliminated but becomes a central constructive principle: quantum correlations help to determine the geometry of spacetime.

\subsubsection{Redundancy in 't~Hooft and in QEC and ER=EPR}

Despite these differences, both 't~Hooft and modern approaches to holography agree on the central role of redundancy. In the case of 't~Hooft, the redundancy lies in the overcompleteness of the bulk description: apparently distinct bulk degrees of freedom are so strongly constrained by the boundary that they amount to different representations of the same boundary degrees of freedom. There are no more independent degrees of freedom in the bulk than can be encoded on the boundary surface. As we have mentioned, this overcompleteness has a counterpart in holographic QEC: in the entanglement wedge picture, more fine-grained bulk localisation (for instance, to $EW(A \cup B)$ rather than $EW(A) \cup EW(B)$) is an overcomplete way of packaging the same boundary information.

However, in QEC-based approaches the primary redundancy is in the \emph{encoding} of gravitational bulk quantities in the boundary theory. There are multiple equally valid boundary candidates for a given bulk observable, and we do not need to privilege one of them: the bulk is encoded redundantly in the boundary code, across many distinct boundary subregions. From the boundary point of view, this redundancy is what enables the recovery of bulk information even in the presence of erasures or inaccessible regions, and underpins the robust emergent bulk description in terms of logical qubits.

In this way, the 't~Hooft and QEC/ER=EPR approaches, despite their differences in scope and interpretation, share conceptual commonalities. Both treat the structure of information---its encoding, redundancy, distribution, and correlations---as central to the emergence of bulk spacetime and to understanding locality, and more broadly the appearance of familiar local physics, from a fundamentally holographic description. Where they differ is in (a) where the redundancy sits (primarily in the bulk, or in the boundary encoding), (b) how the redundancy is interpreted (as infinite correlation or as error-protecting redundancy), and (c) how the underlying microphysics is cashed out: for 't~Hooft, information is ultimately classical and deterministic, implemented in cellular automata; for AdS-CFT, QEC, and ER=EPR, information is intrinsically quantum, and no deterministic underpinning is sought for quantum probabilities and entanglement. Our endorsement in this paper is of these information-theoretic roles of encoding, reundancy, distribution, and correlations, that comprise (iii), rather than of ’t~Hooft’s specific proposal for an underlying deterministic completion of quantum mechanics.

\section{Two conceptions of `information' in holographic setups}\label{notionofinfo}

Having discussed the three aspects of the role of information, (i)-(iii), we now ask how they bear on the kinds of conception of information that are useful in holographic setups. Our aim here is not to give a verdict on which account of information is correct (for a discussion of some options, see Section \ref{intro}, especially footnote \ref{inforefs}), but to indicate how the three roles that we have discussed constrain which accounts of information are suitable in this setting.

First of all, note that the talk about `information' here cannot simply be eliminated in favour of more familiar notions like theories, state-spaces, or dynamics. For, especially under role (iii), some of the discussions are distinctly framed in the language of quantum information theory, and its specific jargon of informational redundancy, encoding, error-correction, logical and physical qubits, entanglement structure, etc. This means that the work we have discussed comes with a characteristic way of formulating physical theories, in informational terms. 

The main point for any future accounts of information in this setting is that the three roles we have identified go beyond familiar discussions in black hole thermodynamics and Shannon-style information theory, which often restrict `information' to epistemic or purely syntactic roles (e.g.~in terms of data compression and uncertainty). 

The conception of information suggested under role (i) is not epistemic: it is not about the knowledge status of agents. Rather, the amount of information is measured by a quantum statistical mechanical entropy, which counts physical states regardless of what specific agents know about these states.

The notion of information suggested under roles (ii) and (iii) also goes well beyond Shannon-style, `engineering', notions of information as uninterpreted data. We thereby distinguish between maximal and intermediate conceptions of information:\\
\\
{\it Maximal conceptions of information.} Under role (ii), to say that dual theories `contain the same information' is to say that they are theoretically equivalent, i.e.~that they are formally equivalent (i.e.~isomorphic) and that they have the same interpretation. For, although questions of encoding, compressibility and redundancy are indeed pertinent (especially under role (iii)), 't Hooft's presumption is an ontological one: the states on the boundary {\it are} the system's states, regardless of whether they are formulated on the boundary or in the bulk. More generally, the standard assumption in discussions of theoretical equivalence and emergence in holographic settings is that dualities relate {\it interpreted} theories. Thus the use of `same information content' in this context commits us to a maximal conception of information: namely, one that is concerned with the ontology of the theory, and not only with its syntactic, epistemic or semantic content.\footnote{It is a standard requirement of theoretically equivalent theory formulations that they make the same ontological claims. See for example Sklar (1982:~p.~90) and Glymour (1975:~p.~237). For the requirement in the context of dualities, see De Haro and Butterfield (2025:~pp.~82, 418) and Read and M\o ller-Nielsen (2020). Also in discussions of emergence, a distinction is routinely made between epistemic and ontological emergence (De Haro and Butterfield, 2025:~p.~500).} 
This usage of `same information content' as a gloss for theoretical equivalence thus requires a maximal conception of information.

There is also a second, intermediate, type of conception of information that is not committed to theoretical equivalence and still goes well beyond Shannon-style accounts:\\
\\
{\it Intermediate conceptions of information.} Under role (iii), holography is not just about encoding uninterpreted, unstructured, data. Reformulated in information-theoretic terms, the holographic duality encodes detailed information about states, quantities, and dynamics in quantum error-correcting structures, often modelled by quantum circuit or tensor-network representations. Also when there is not a duality, but rather a relation of emergence, what is being encoded by the holographic map is not just a number of qubits, but also the logical structures in which these qubits operate---so that logical qubits emerge by coarse-graining physical qubits.

Unlike the maximal conceptions of information, an intermediate conception does not require the theoretical equivalence of duals: it reformulates the duality in quantum information-theoretic terms. But this does not mean that such a conception is independent of interpretation. For the structural equivalence of two theories required by a duality is a strong condition: the two theories must share representational capacities. In other words, they share isomorphic state-spaces, sets of quantities, and dynamics. Much of theoretical physics in practice consists precisely in identifying the structures in terms of which a theory may be described. We therefore call such conceptions of information `intermediate': they go beyond both epistemic notions of information and the conception of information as `purely uninterpreted data' in the sense of Shannon. Even prior to the fixing of a full interpretation, physical theories possess rich mathematical and conceptual structures that constrain their admissible interpretations.\footnote{In this sense, one sometimes speaks of the `proto-interpretation' of a theory. For a discussion of how uninterpreted theories constrain the theory's descriptive capacities, see De Haro and Cinti (2026:~p.~10).}
Such a conception need not directly concern the theory's ontology, but it nonetheless involves a structural semantics.

\section{Conclusion}\label{conclusion}

In this paper we have argued that ’t Hooft’s early formulation of the holographic principle illustrates how information functions as a substantive guiding principle in quantum gravity practice. We distinguished three interrelated aspects of this role. (i) Entropy bounds on the amount of information that can be stored in a region function as theory selection criteria: they motivate the holographic principle and constrain the admissible quantum gravity theories, including in recent Swampland arguments that rule out certain effective field theories. (ii) Holography functions, for 't Hooft, as a requirement for theoretical equivalence and its assessment: ’t Hooft’s viability criterion requires that the physical content of any bulk description be fully captured by boundary degrees of freedom, even as he privileges a more fundamental, lower dimensional boundary dynamics that he hopes can be rendered deterministic. As such, this mirrors a similar tension in the recent literature on duality and emergence of spacetime. (iii) We highlighted the distribution of information: ’t Hooft’s emphasis on infinite bulk correlations and bulk redundancy resonates with contemporary accounts in which bulk spacetime and locality emerge from the encoding, redundancy, and entanglement structure of boundary degrees of freedom, as in holographic quantum error-correction and ER=EPR.

These three aspects illustrate how, in holography and quantum gravity, information does not function as either an agent-relative or a purely formal notion, but rather as a substantive guide to theory construction and interpretation. This approach has proven to be remarkably fruitful in quantum gravity. Even if one remains suitably cautious about reifying `information', one cannot ignore its methodological import and its role as an interpretation-guiding principle in holography and quantum gravity. Purely epistemic or syntactic notions of information cannot fulfil these roles. The maximal and intermediate conceptions of information that we have distinguished help to spell out how these richer roles outrun purely Shannon-style notions, and thereby help to articulate the sense in which the language of information, in holographic quantum gravity, is not a `fa\c{c}on de parler', but a substantive way of formulating the theory’s structure and guiding its interpretation. It would therefore be appropriate to develop more systematic philosophical accounts of how information can function as a guiding and interpretative principle in physical theories.

\section*{Acknowledgements}
\addcontentsline{toc}{section}{Acknowledgements}

SDH and EC's work is supported by the John Templeton Foundation grant 63670. SDH also thanks Jesus College, Oxford, for their hospitality and support during the writing of this paper. EC's is supported also by the SNSF project {\it Space, time, and causation in quantum gravity}.

\section*{References}
\addcontentsline{toc}{section}{References}

\small

Adriaans, P.~(2023). `Information'. {\it Stanford Encyclopedia of Philosophy.}\\ https://plato.stanford.edu/entries/information. 

\ \\Adriaans, P.~and van Benthem, J.~(2008). {\it Philosophy of Information}. Handbook of the Philosophy of Science, Volume 8. Amsterdam: Elsevier.

\ \\Aharony, O., Gubser, S.S., Maldacena, J.M., Ooguri, H.,~and~Oz, Y.~(2000). `Large $N$ Field Theories, String Theory and Gravity'. {\it Physics Reports}, 323, pp.~183-386.

\ \\Almheiri, A., Dong, X. and Harlow, D.~(2015). `Bulk locality and quantum error correction in AdS/CFT.' \textit{Journal of High Energy Physics}, 163 (2015).

\ \\Ammon, M.~and Erdmenger, J.~(2015). {\it Gauge/Gravity Duality}. Cambridge: Cambridge University Press.

\ \\Arkani-Hamed, N., L.~Motl, A.~Nicolis and C.~Vafa (2007). `The string landscape, black holes and gravity as the weakest force'. {\it Journal of High-Energy Physics}, 6, 60, pp.~1-16.

\ \\ Bain, Jonathan (2020). Spacetime as a quantum error-correcting code? \textit{Studies in History and Philosophy of Science Part B: Studies in History and Philosophy of Modern Physic},s 71 (C), pp.~26-36.

\ \\Bekenstein, J.~D.~(1973). `Black Holes and Entropy'. {\it Physical Review D}, 7 (8), pp.~2333-2346.

\ \\Bekenstein, J.~D.~(1981). `Universal upper bound on the entropy-to-energy ratio for bounded systems'. {\it Physical Review D}, 23 (2), pp.~287-298.

\ \\Bousso, R.~(2002). `The holographic principle'. {\it Review of Modern Physics}, 74, pp.~825-874. 

\ \\Bousso, R., Freivogel, B., Leichenauer, S., Rosenhaus, V., and Zukowski, C.~(2013). `Null
geodesics, local cft operators, and ads/cft for subregions'. {\it Phys. Rev. D} 88, p.~064057.

\ \\Corbeel, A.~and De Haro, S.~(2026). `Black Holes are about Quantum Information'. https://philsci-archive.pitt.edu/id/eprint/27365.

\ \\Cover, T.~M.~and Thomas, J.~A.~(2006) [1999]. {\it Elements of Information Theory}, Second Edition. Hoboken: Wiley.

\ \\De Haro, S.~(2017). `Dualities and emergent gravity: Gauge/gravity duality'. {\it Studies in History and Philosohpy of Modern Physics}, 59, pp.~109-125.

\ \\De Haro, S.~(2021). `Theoretical equivalence and duality'. {\it Synthese}, 198, pp.~5139–5177.

\ \\De Haro, S.~and J.~Butterfield (2025). {\it The Philosophy and Physics of Duality}. Oxford: Oxford University Press.

\ \\De Haro, S.~and Cinti, E.~(2026). {\it Dualities in Physics}. Cambridge: Cambridge University Press.

\ \\De Haro, S., Mayerson, D.~R.~and Butterfield, J.~N.~(2016). `Conceptual Aspects of Gauge/Gravity Duality'. {\it Foundations of Physics} 46, pp.~1381–1425.

\ \\Dieks, D., van Dongen, J.~and De Haro, S.~(2015). `Emergence in holographic scenarios for gravity\. {\it Studies in History and Philosophy of Modern Physics}, 52 (B), pp.~203-216.

\ \\Dong, X., Harlow, D. and Wall, A. C. (2016). ‘Reconstruction of Bulk Operators within the Entanglement Wedge in Gauge-Gravity Duality’. \textit{Physical Review Letters}, 117(2), p. 021601.

\ \\Dougherty, J.~and C.~Callender (2017). `Black hole thermodynamics: More than an analogy?'' 

\ \\Dretske, F.~I.~(1981). {\it Knowledge and the Flow of Information}. Cambridge, MA: MIT Press.

\ \\Dvali, G.~(2010). `Black holes and large $N$ species solution to the hierarchy problem'. {\it Fortschritte der Physik}, 58 (6), pp.~528--536.

\ \\Dvali, G.~and Redi, M.~(2008). `Black hole bound on the number of species and quantum gravity at CERN LHC'. {\it Physical Review} D, 77, 045047, pp.~1--8.

\ \\Engelhardt, N., Liu, H.~(2024). `Algebraic ER=EPR and complexity transfer'. \textit{Journal of High Energy Physics}, 13 (2024). 

\ \\Floridi, L.~(2011). {\it The Philosophy of Information}. Oxford: Oxford University Press.

\ \\ Goldstein, S.~(2025) "Bohmian Mechanics", {\it The Stanford Encyclopedia of Philosophy}, Edward N. Zalta \& Uri Nodelman (eds.), https://plato.stanford.edu/archives/fall2025/entries/qm-bohm/.

\ \\Glymour, C.~(1977). ‘The epistemology of geometry’. No$\hat{\mbox u}$s, pp.~227--251.

\ \\Hamada, Y., Montero, M., Vafa, C.~and Valenzuela, I.~(2022). `Finiteness and the swampland'. {\it Journal of Mathematical Physics A: Mathematical and Theoretical}, 55, 224005, pp.~1-25.

\ \\Hubeny, V. E., M. Rangamani, and T. Takayanagi (2007). `A covariant holographic entanglement entropy proposal'. \textit{Journal of High Energy Physics}, 07 (2007).

\ \\Kolmogorov, A.~N.~(1965). `Three approaches to the quantitative definition of information'. {\it Problems of information transmission}, 1 (1), pp.~1-7.

\ \\Maldacena, J. and Susskind, L. (2013). ‘Cool horizons for entangled black holes'. \textit{Fortschritte der Physik}, 61(9), pp. 781–811.

\ \\Ooguri, H.~and Vafa, C.~(2007). `On the geometry of the string landscape and the swampland'. {\it Nuclear Physics B}, 766, pp.~21-33.

\ \\Palmer, T.~(2026). `Rational Quantum Mechanics: In Support of Ontological Bases and Superdeterminism' {\it This Volume}.

\ \\Read, J.~and T.~M\o ller-Nielsen (2020). `Motivating dualities\ . {\it Synthese} 197, pp.~263–291.

\ \\Sequoiah-Grayson, S.~and Floridi, L.~(2022). `Semantic Conceptions of Information'. {\it Stanford Encyclopedia of Philosophy}. https://plato.stanford.edu/entries/information-semantic.

\ \\Schumacher, B.~(1995). Quantum coding. {\it Physical Review A}, 51, pp.~2738–2747.

\ \\Shannon, C.~E.~(1948). `A Mathematical Theory of Communication'. {\it The Bell System Technical Journal}, 27, pp.~379-423, 623-656.

\ \\Sklar, L.~(1982). `Saving the Noumena’. {\it Philosophical Topics}, 13 (1), pp.~89--110.

\ \\Susskind, L.~(1995). `The world as a hologram'. {\it Journal of Mathematical Physics}, 36 (11), pp.~6377-6396.

\ \\Susskind, L.~and E.~Witten (1998). `The Holographic Bound in Anti-de Sitter Space'. ArXiv: hep-th/9805114.

\ \\'t Hooft, G.~(1991). `The black hole horizon as a quantum surface'. {\it Physica Scripta}, T36, pp.~247-252.

\ \\'t Hooft, G.~(1994). `Dimensional reduction in quantum gravity'. In: {\it Salamfestschrift. A collection of talks from the conference on highlights of particle and condensed matter physics}. Ali, A., Ellis, J., Randjbar Daemi, S. (Eds.), pp.~284–296. Singapore: World Scientific. ArXiv: gr-qc/9310026.

\ \\'t Hooft, G.~(2001). `The holographic principle'. In: {\it Basics and Highlights in Fundamental Physics}, A.~Zichichi (Ed.), pp.~72-100. Singapore: World Scientific Publishing. ArXiv: hep-th/0003004.

\ \\Timpson, C.~G.~(2013). {\it Quantum Information Theory and the Foundations of Quantum Mechanics}. Oxford: Oxford University Press.

\ \\Van Fraassen, B.~C.~(1991). {\it Quantum mechanics: An empiricist view}. Oxford University Press.

\ \\Verlinde, E.~(2011). `On the origin of gravity and the laws of Newton'. {\it Journal of High Energy Physics}, no. 4: 1-27.

\ \\Verlinde, E.~(2017). `Emergent gravity and the dark universe'. {\it SciPost Physics 2}, no. 3: 016.

\ \\Wall, A. C.~(2014). ‘Maximin Surfaces, and the Strong Subadditivity of the Covariant Holographic Entanglement Entropy’. \textit{Classical and Quantum Gravity}, 31(22), p. 225007.

\ \\Wallace, D.~(2008). `The quantum measurement problem: State of play'. In: {\it The Ashgate Companion to Contemporary Philosophy of Physics}, D.~Rickles (Ed.), Ashgate.

\ \\Wallace, D.~(2018). `The case for black hole thermodynamics part i: Phenomenological thermodynamics'. {\it Studies in History and Philosophy of Modern Physics} 64, pp.~52–67. 

\ \\Wallace, D.~(2019). `The case for black hole thermodynamics part ii: statistical mechanics'. {\it Studies in History and Philosophy of Modern Physics} 66, pp.~103–117.

\ \\Wheeler, J.~A.~(1989). `Information, physics, quantum: The search for links'. In: {\it Proceedings III International Symposium on Foundations of Quantum Mechanics}, pp.~354–358.

\ \\W\"uthrich (2017). `Are black holes about information?' In: {\it Epistemology of Fundamental Physics: Why Trust a Theory?}, Dardasti, R., Dawid, R.~and Th\'ebault, K.~(Eds.). Cambridge: Cambridge University Press.

\end{document}